\documentclass[aps,twocolumn,superscriptaddress,showpacs]{revtex4-1}

\usepackage[utf8x]{inputenc} 
\usepackage[T1]{fontenc}
\usepackage[english,american]{babel}

\usepackage{lmodern}

\usepackage{textcomp,gensymb}
\usepackage{amssymb,amsmath}
\usepackage{units}
\usepackage{bm}
\usepackage{isotope}

\usepackage{graphicx}

\usepackage{hyperref}
\hypersetup{
  pdfstartview={FitH},
  pdftitle={},
  pdfauthor={Ian D. Leroux}}

\ifpdf
\fi

\begin{document}

\title{Sub-Micron Positioning of Trapped Ions with Respect to the Absolute Center of a Standing Wave Cavity Field}

\newcommand*{\IFAAU}{
QUANTOP, Danish National Research Center for Quantum Optics, Department of Physics and Astronomy, Aarhus University, DK-8000 Aarhus C, Denmark}
\author{Rasmus B. Linnet} \affiliation{\IFAAU}
\author{Ian D. Leroux} \affiliation{\IFAAU} \affiliation{Physikalisch-Technische Bundesanstalt, Bundesallee 100, 38116 Braunschweig, Germany}
\author{Aurélien Dantan} \affiliation{\IFAAU}
\author{Michael Drewsen} \email{drewsen@phys.au.dk} \affiliation{\IFAAU}
\date{\today}


\begin{abstract}
We demonstrate that it is possible, with sub-micron precision, to locate the absolute center of a Fabry-P\'{e}rot resonator oriented along the rf-field-free axis of a linear Paul trap through the application of two simultaneously resonating optical fields. In particular, we apply a probe field, which is near-resonant with an electronic transition of trapped ions, simultaneously with an off-resonant strong field acting as a periodic AC Stark-shifting potential. Through the resulting spatially modulated fluorescence signal we can find the cavity center of an 11.7 mm-long symmetric Fabry-P\'{e}rot cavity with a precision of $\pm 135$ nm, which is smaller than the periodicity of the individual standing wave fields. This can e.g. be used to position the minimum of the axial trap potential with respect to the center of the cavity at any location along the cavity mode. 
\end{abstract}

\maketitle

\section{Introduction}
\label{intro}
For more than half a century, it has been possible to confine charged particles by radiofrequency (rf) quadrupole fields~\cite{Paul1990}. Such rf (or Paul) traps rely on the strong direct interaction of the charge of the particles with the electrical field, whereby trap depths of up to hundreds of eV (i.e., trapping temperatures of the order of a million Kelvin) can easily be achieved for atomic ions. This makes it generally easy to trap ions by e.g. introducing a buffer gas at room temperature to provide a friction force which damps the ion motion. Conversely, trapping of neutral atomic species by electric fields has to rely on the interaction of an induced electric dipole moment with the field itself. Although such interactions can be enhanced significantly by choosing electric fields oscillating close to an electronic transition frequency of the atom, the depth of such traps typically do not exceed 100 mK~\cite{Grimm2000}. Hence, in order to achieve trapping, laser cooling has to be implemented. In the past decades, cooling and trapping of neutral atoms in various dipole-induced lattice trap configurations has led to the studies of a wealth of physics phenomena, by mimicking for instance idealized solid-state physics scenarios, such as e.g. Bloch-oscillations~\cite{BenDahan1996} and the superfluid-MOT insulating transition~\cite{Greiner2002}.

Laser cooling of atomic ions~\cite{Wineland1975} in Paul~\cite{Leibfried2003} and Penning~\cite{Brown1986} traps has as well facilitated the studies of a large variety of physics phenomena ranging from pure plasma physics~\cite{Dubin1999}, non-linear physics~\cite{Blumel1988,Hoffnagle1988,Blumel1989,Brewer1990} to quantum physics~\cite{Leibfried2003}, including quantum information processing~\cite{Blatt2008}, and has even opened a whole new field of cold molecular ion-based research~\cite{Willitsch2008}. However, electric field-induced dipolar forces have only recently been applied to trap, or alter the trapping conditions of, atomic ions~\cite{Schneider2010,Enderlein2012,Linnet2012,Karpa2013}. The interest here has been partly to demonstrate trapping of a single ion with localized fields in order to e.g. enable coherent interactions between atoms and ions without perturbation induced by trapping fields~\cite{Cormick2011}, partly to superpose a steep periodic potential to a shallow rf trap potential with the aim of studying structural~\cite{Drewsen2012,Horak2012,Cormick2012} and dynamical phase transitions (e.g. Coulomb-Frenkel-Kontorova model~\cite{GarciaMata2007,Benassi2011,Pruttivarasin2011,Schneider2012review}), as well as enhancing the coupling strength between ions and cavity photons with quantum memory~\cite{Herskind2009,Albert2011} and photon counter~\cite{Clausen2013} applications in mind. With respect to the latter applications, standing wave fields generated in a Fabry-P\'{e}rot cavity are of special interest, since they provide a means to achieve a well-controlled spatial phase between a localizing field mode and an interrogation field mode at the position of the ions~\cite{Linnet2012}. Fine probing of the longitudinal cavity field spatial structure has been performed with single ions~\cite{Guthohrlein2001,Mundt2002}; however, the application of a single standing-wave field does not allow for the absolute positioning discussed here. Due to the boundary conditions for the fields in a Fabry-P\'{e}rot cavity, all modes of the same parity (even or odd) have overlapping nodes and antinodes at the center (waist) of the cavity, i.e. an in-phase relation is imposed at the center, regardless of the frequency of the modes~\cite{note}. Similarly, an out-of-phase relation can here be obtained by combining field modes with even and odd number of nodal planes. With an exact knowledge of the center position of an applied optical cavity one can hence deterministically switch between having the trapped ions at a node or anti-node of the potential, as has e.g. been exploited in some of the experiments reported in~\cite{Linnet2012}. For large ensembles of ions, it could also be interesting to be perform this with respect to a corrugated super-lattice created through the interference of two cavity modes.

In this paper, we demonstrate a simple method to determine the center of a near-confocal symmetric optical Fabry-P\'{e}rot cavity having its rotational symmetry axis aligned with the rf nodal line of a linear rf trap~\cite{Herskind2009JPB}, by utilizing an ion Coulomb crystal as an imaging medium. In the following section (Sec. \ref{CenterCavTheory}), we first consider the idealized case of a two-level atom interacting with two cavity modes, a probe field oscillating at a frequency close to the two-level system resonance frequency, and an off-resonant lattice field providing a periodic AC Stark potential. In Sec. \ref{Exp_setup}, we describe the essential parts of the experimental setup used in our investigations with Ca$^+$ ions. This section is followed by a description of the experimental procedure and obtained results (Sec. \ref{sec:findcentre}). In Sec. \ref{Future}, we discuss some future prospects of this method for multi-cavity mode operation with cold ions and atoms, before concluding in Sec. \ref{Conclusion}.

\section{Simple two-level description}
\label{CenterCavTheory}
A full description of the applied method for centering the trapping potential with respect to the center of a standing wave light field of a Fabry-P\'{e}rot cavity would have to be based on an ensemble of multi-state systems with the effects of the applied magnetic and electrical fields on the individual states included. However, in order to provide a clear physical picture, in this section, we provide only a description for an ensembles of two-level atoms.

\begin{figure}
  \centering \includegraphics[width=1\linewidth]{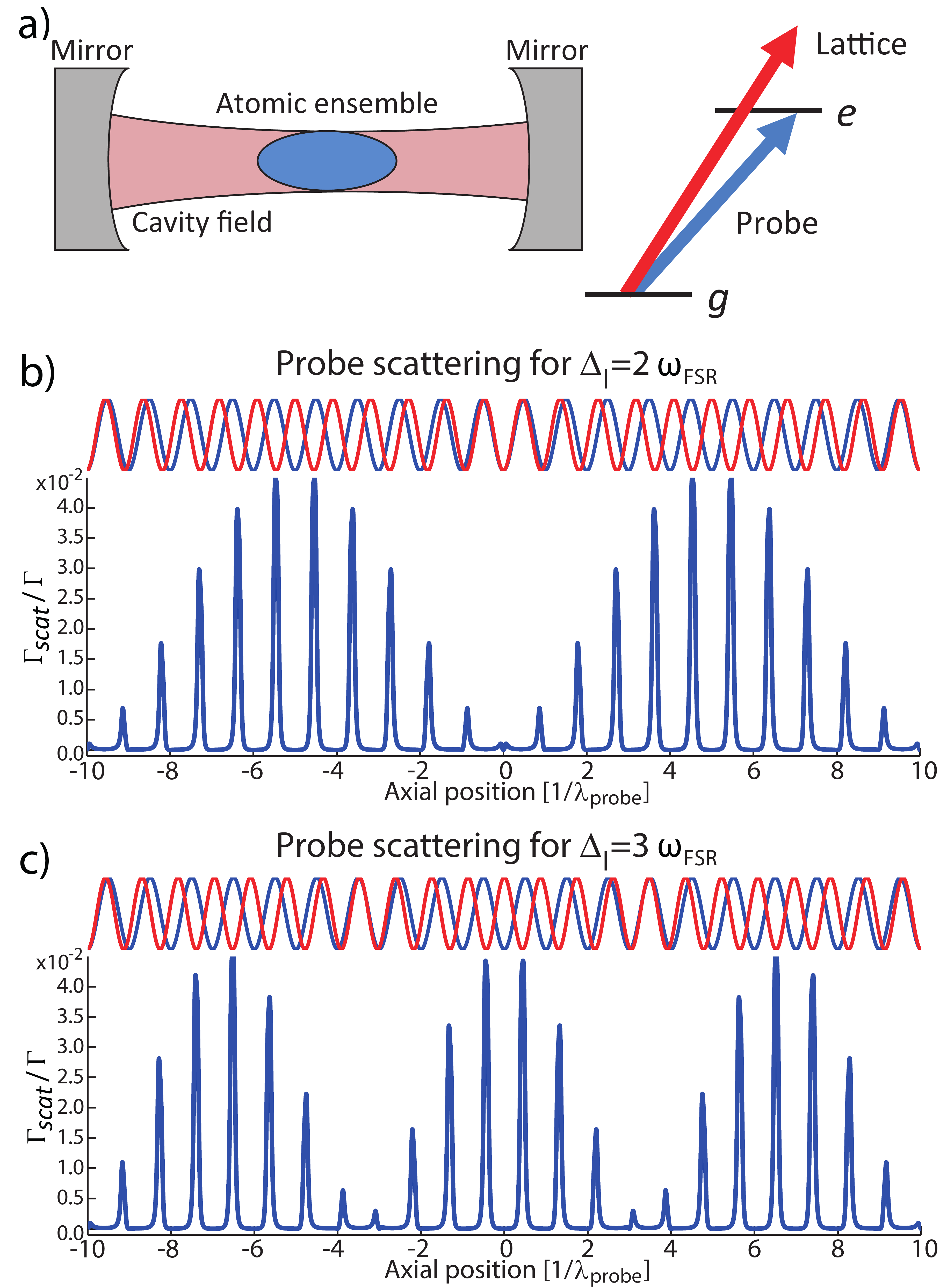}
  \caption{a) Schematics of the considered atomic two-level system with energy levels and applied fields. In b) and c) the parameters $n_p=20$ and $n_l=22,23$ have been used, together with $L=10$, $\Gamma=1$, $\Delta_s=10$ and $s_0=0.1$, in order to illustrate the effect. The real values of $n_p$ and $n_l$ in the experiment are around $8.49 \times 10^{4}$. The lower part of the figures (blue peak structure with a wide envelope) shows the variation of the probe scattering, $\Gamma_{\textrm{scat}}$, along the cavity axis, for a lattice detuning of b) 2$\omega_{FSR}$ and c) 3$\omega_{FSR}$. The top part of the figures (sinusoidal curves) shows the probe standing wave (blue) and the effective Stark shifted detuning $\Delta(z)$ (red) along the cavity axis; the two sinusoidal curves in this part of the figure have been rescaled to the same amplitude in order to better illustrate the spatial beating.}
  \label{fig:CenterCavResults}
\end{figure}

More specifically, we consider two-level atoms with a\\ ground state \textit{g} and an excited state \textit{e} positioned inside a near-confocal, symmetric optical Fabry-P\'{e}rot cavity (see Fig.~\ref{fig:CenterCavResults}a). The atoms are trapped by an external mechanism which keeps them confined within the cavity mode-volume. A so-called lattice field, which is far-detuned from the atomic resonance, but on resonance with a longitudinal mode of the cavity, is applied. By keeping one of the cavity modes resonant with the \textit{g}-\textit{e} transition this leads to a lattice detuning of a whole number of cavity free-spectral-ranges, $\omega_{FSR}$ from the two-level resonance. The effect of the lattice field is to induce a spatially modulated AC Stark shift of the atomic transition given by:
\begin{equation}
  \Delta_S(z) = \Delta_S \sin^2(k_{l} z)
  \label{eq:S_z}
\end{equation}
where $\Delta_S$ is the maximum Stark shift, $k_{l}=(n_{l} \pi) / L$ the lattice field wavenumber and $n_{l}$ the longitudinal mode number of the lattice field.

The effect of this lattice field is monitored by a near-resonant intercavity probe field detuned $\Delta_{p}$ with respect to the bare two-level transition frequency. The spatially dependent photon scattering rate of this probe field is given by:
\begin{equation}
  \Gamma_{\textrm{scat}} = \frac{1}{2} \frac{s(z)}{1+s(z)} \Gamma
  \label{eq:Pie}
\end{equation}
where $s(z)$ is the saturation parameter, defined as:
\begin{equation}
  s(z)=\frac{s_0 \sin^2(k_p z)}{1+\left(2 \Delta(z) / \Gamma\right)^2},
  \label{eq:satpar}
\end{equation}
with $\Gamma$ being the decay rate of the excited state, $s_0=I_0/I_{sat}$ the maximum on-resonance saturation parameter for the probe field, $k_{p}=(n_{p} \pi) / L$ the probe field wavenumber expressed in terms of $n_{l}$, the longitudinal mode number of the lattice field, and the spatially varying effective detuning is given by 

\begin{equation}
  \Delta(z) = \Delta_{p} + \Delta_S \sin^2(k_{l} z)
  \label{eq:delta}
\end{equation}

The effect of the Stark shifting lattice field on the probe photon scattering rate (eq.~\ref{eq:Pie}) is a beating signal arising from the wavenumber difference of the probe and lattice fields. This is illustrated in Fig.~\ref{fig:CenterCavResults}b-c where the parameters have been chosen to illustrate the effect ($n_p = 20$ and $n_l =22,23$). In the experiment the values of $n_p$ and $n_l$ are around $8.49 \times 10^{4}$. In most practical realizations the imaging system is not able to resolve the fine-structured pattern, as the individual lattice sites are typically separated by only a few hundreds of nm. The beating signal, on the other hand, occurs on a much larger length scale (proportional to the inverse of the probe-lattice wavevector difference) and can be resolved using standard imaging techniques, as will be shown below. At the center of the cavity ($z=0$) the beat pattern has an extremum because of the boundary conditions imposed by the mirrors. If the lattice detuning is an even number of free-spectral-ranges (FSRs) the scattering rate is minimised, as the probe and lattice fields overlap so that the transition is shifted away from the probe frequency wherever the probe field is strong.  A lattice detuning by an odd number of FSRs produces a maximum of the scattering rate, as the probe field is strongest where the transition is unshifted. This is also illustrated in Fig.~\ref{fig:CenterCavResults}b-c.

As mentioned earlier, the two-level description cannot be expected to give a precise account of measured scattering rates. However, around the center of the Fabry-Perot cavity, the scattering rate will always vary periodically with a spatial period set by the inverse of the probe-lattice wavevector difference, $\lambda_{beat} \propto 1/(k_{l}-k_{p})$, with a maximum (minimum) for $n_p-n_l$ being odd (even). The length scale of $\lambda_{beat}$ for our experimental parameters is hundreds of $\mu$m.

\section{The experimental setup}
\label{Exp_setup}
\begin{figure}
  \centering
  \includegraphics[width=1\linewidth]{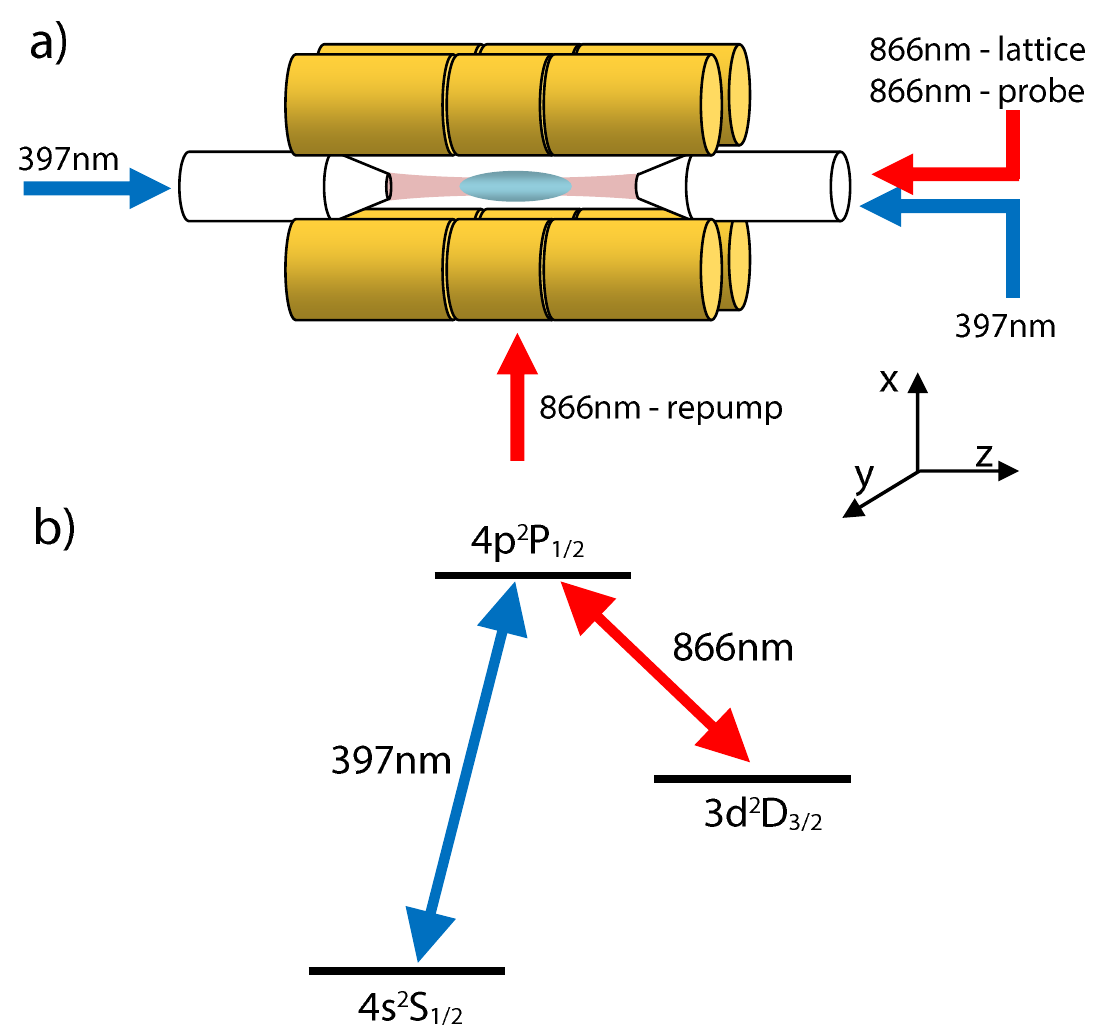}
  \caption{a) Schematic picture of the cavity ion trap and laser fields used in the experiments in top view (see text for details). b) Relevant energy levels for $^{40}$Ca$^+$.}
  \label{fig:setup}
\end{figure}

The setup used in the experiments has been described in detail in~\cite{Herskind2008} and is depicted schematically in Fig.~\ref{fig:setup}. It consists of a symmetrically-driven four-rod linear Paul trap operating at a 4 MHz drive frequency. The trap incorporates a pair of mirrors forming a near-confocal Fabry-P\'{e}rot resonator whose optical axis is aligned with the nodal line of the trap electric fields~\cite{Herskind2009JPB}. The cavity has a length of 11.7 mm, corresponding to a free-spectral-range $\omega_{\mathrm{FSR}} = 2 \pi \times 12.7$ GHz. It has a finesse of $\sim 3000$, and a zeroth-order mode waist radius of 37 $\mu$m for light at 866 nm. The trap is loaded with $^{40}$Ca$^+$ ions, whose relevant energy levels and transitions are depicted in Fig.~\ref{fig:setup}b). The $S_{1/2}-P_{1/2}$ transition at 397 nm is used for Doppler cooling and imaging, while the $D_{3/2}-P_{1/2}$ transition at 866 nm is used either for repumping ions shelved in the metastable $D_{3/2}$ state during cooling or for interactions with the cavity light. In Fig.~\ref{fig:setup}a) the applied laser fields are also sketched. A bias magnetic field of 1 Gauss is applied in the $y$-direction. The 397 nm cooling light is applied along the axial ($z$) direction (contrapropagating beams with opposite circular polarizations) and the 866 nm repump light along the radial ($x$) direction with linear polarization along the $z$-direction. The ions can be imaged by collecting the 397 nm fluorescence onto an intensified CCD camera with line-of-sight along the $y$-direction (not shown). Two different 866 nm laser fields with circular polarization can be coupled into the cavity: the resonant \textit{probe} field and the off-resonant \textit{lattice} field. The cavity length is stabilized using an additional off-resonant laser (not shown), which is locked to an external reference cavity and which has negligible influence on both the internal and external states of the ions.

\section{Finding the center of the optical cavity}
\label{sec:findcentre}

\begin{figure*}
  \centering
  \includegraphics[width=0.95\textwidth]{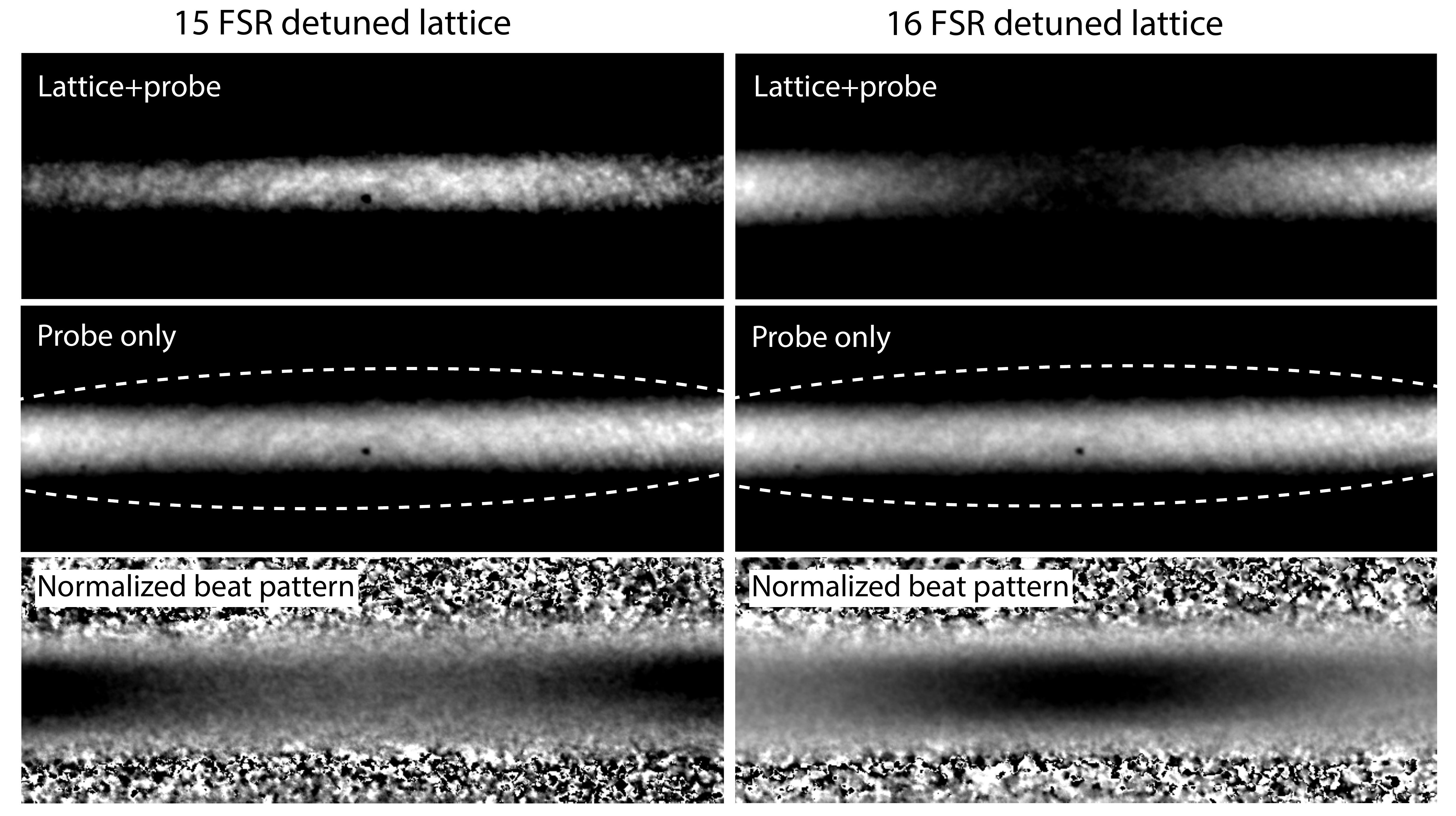}
  \caption{Experimental images ($224 \times 640$ pixel) of a $^{40}$Ca$^+$ ion Coulomb crystal with $\approx 6000$ ions (length 650 $\mu$m, diameter 150 $\mu$m, density $~\sim 6 \times 10^8$ cm$^{-3}$). As repumping is only performed by the intracavity fields, only the central part of the crystal, contained in the cavity modevolume, is visible. For reference, an outline of the actual crystal shape is shown by the dashed ellipsoids in the middle pictures. The top figure shows the ion scattering signal when applying both the lattice and probe fields. Two situations are shown corresponding to a lattice field detuned by $+15$FSR and $+16$FSR, respectively; a clear beating signal is observed, with a maximum at the cavity center in the first case and a minimum in the second. After background subtraction, making the pixel-to-pixel ratio of the top and middle images gives the bottom image, thus isolating the fluorescence beat pattern due to the lattice field. The images are obtained from $60\times 200$ ms exposures.}
  \label{fig:CenterCavCrystals}
\end{figure*}

In a recent study~\cite{Linnet2012} it was shown that an ion could be pinned by the intracavity standing wave lattice field in the described cavity ion trap. One important issue in such experiments is to ensure that the overlap between the probe and the lattice fields is well-controlled at the ion position. This means that the ion has to be precisely located at the center of the cavity to obtain reproducible situations for all lattice detunings. To locate this center we trap a large ion Coulomb crystal of $^{40}$Ca$^+$ ions and inject into the cavity both the resonant probe field (see Fig.~\ref{fig:setup}) and the lattice field detuned by an odd or even integer number of FSRs away from the atomic resonance.

Due to the lattice field Stark shifts of the energy levels of the $P_{1/2}$ and $D_{3/2}$ states, the scattering rate of the probe field becomes spatially modulated, as discussed in Sec.~\ref{CenterCavTheory}. In regions where the high (low) intensity of the lattice overlaps with the high intensity of the probe this shift is largest (smallest) and the probe scattering is suppressed (unchanged). As mentioned earlier, our imaging resolution ($\sim \mu$m) does not allow us to resolve standing waves at the single lattice site scale, but the resulting beating signal is observable on a larger scale.

In the experiment, an ion Coulomb crystal containing $\sim 6000$ $^{40}$Ca$^+$-ions (length 650 $\mu$m, diameter 150 $\mu$m, density $~\sim 6 \times 10^8$ cm$^{-3}$) is trapped and Doppler-cooled using the axial 397 nm cooling beams and the 866 nm radial repumper. The bias magnetic field applied in the $y$-direction ensures that the circularly polarized cavity fields address all four Zeeman sublevels of the $D_{3/2}$ state. The cavity length is chosen so that, in the absence of a lattice field, the probe field can be simultaneously resonant with the $D_{3/2}-P_{1/2}$ transition and with a cavity mode.

When the cavity fields are applied, the 866 nm side repumper is blocked, so as to perform all repumping through the cavity only. The amplitude of the spatially varying saturation parameter for the probe is $s_{max} \approx 4$ (eq.~\ref{eq:satpar}) while that of the lattice is at least $1000$ times smaller, in order to realize a situation where scattering essentially comes from the probe field and the lattice field only produces a spatially dependent Stark shift. We set the probe laser on resonance with $\Delta_p = 2 \pi \times (0 \pm 2)$ MHz and the lattice Stark shift depends on the detuning of the lattice field and is in the range $\Delta_s = 2 \pi \times (3-9)$ MHz. The 397 nm laser is red-detuned by 40 MHz, so that the lattice Stark shift does not appreciably affect its scattering rate. The observed fluorescence modulation is therefore dominated by the variation in the repumping rate out of the $D_{3/2}$ state. 

\begin{figure}
  \centering
  \includegraphics[width=1\linewidth]{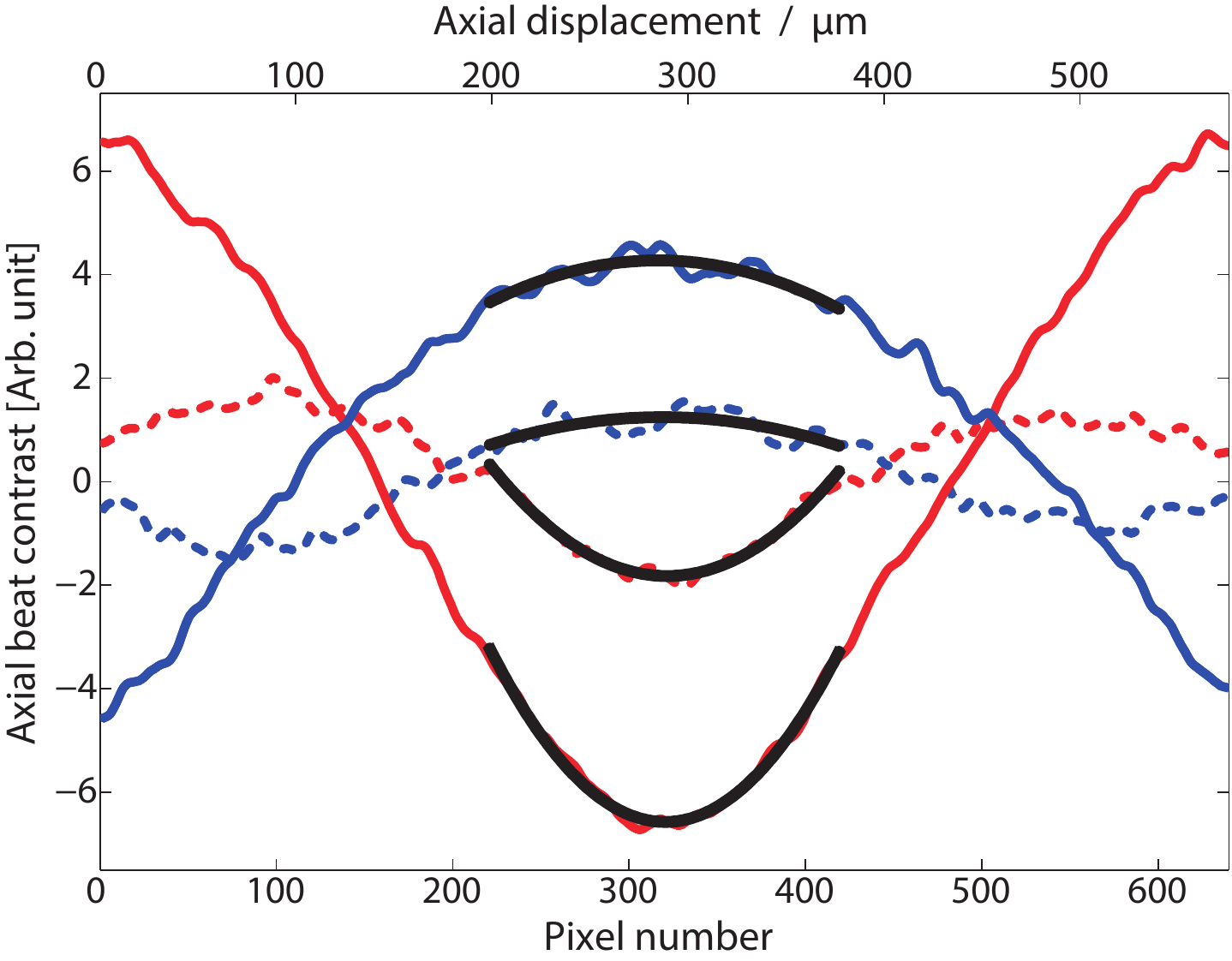}
  \caption{Beating signals along the axis of a Coulomb crystal when both the lattice and probe fields are injected into the cavity. Depending on the lattice detuning from the probe the characteristics of the beating change and this is shown for four different detunings: 15 FSR (full blue), 16 FSR (full red), 27 FSR (dashed blue), 28 FSR (dashed red). Detuning by an even number of FSRs (here, 16 and 28) results in a minimum scattering at the cavity center, while detuning by an odd number of FSRs (here 15 and 27) produces a maximum. At the center of the cavity maxima and minima line up, as expected. The black parabolic fit is performed on $\pm100$ pixels around the center.}
  \label{fig:Exp_results_cav_center}
\end{figure}

In Fig.~\ref{fig:CenterCavCrystals} projection images of the crystal are shown for lattice detunings of $+15$FSR and $+16$FSR. The cavity fundamental mode is clearly visible as only the ions in the crystals contained within the cavity mode volume participate in the cooling cycle and sympathetically cool the ions outside the cavity mode volume. The top figures show images obtained when applying both the lattice and probe fields, while the middle figures show images with the probe field only. After background subtraction, making the ratio of the image with and without the lattice allows for correcting for inhomogeneities of the imaging system and thus isolates the fluorescence modulation pattern due to the lattice. A running average is applied across the beat pattern image setting the single pixel value equal to the mean of pixel values in a $10 \times 10$ square around it. In this way unevenness in the crystal structure from e.g. crystal shells will be blurred to ensure a smother beating signal along the axial direction. The central part of the crystal is then isolated and 37 pixels (33 $\mu m$) are summed in the vertical direction. The result is proportional to the fluorescence signal of ions with the same axial position along the cavity. In Fig.~\ref{fig:Exp_results_cav_center} the resulting modulated signals are shown for a lattice field detuned by 15 (blue), 16 (red), 27 (dashed blue) and 28 (dashed red) FSRs, respectively. As mentioned, the experimental system is more complicated than the simple two-level illustrative picture presented in Sec. \ref{CenterCavTheory} and the envelope of the beat pattern is not given by a simple analytical function that we could use as a model for fitting.  Rather than fitting the full beat pattern to a numerical model, we fit the points around the center of the cavity with a parabolic function in order to establish the exact cavity center position which is all we care about for this purpose.

Beat patterns with only one or a few cycles within the length of the cavity give unambiguous position information, but do not offer much position resolution because of their coarse spatial structure.  Beat patterns with many cycles, obtained using a far-detuned lattice, provide fine resolution but do not distinguish between positions separated by the beat period. By combining images taken for different lattice detunings, one can obtain more precise location information anywhere in the cavity without losing track of the overall position. For example, for the set of beat patterns shown in Fig.~\ref{fig:Exp_results_cav_center}, all maxima and minima line up, as they must, at the cavity center. Since e.g. $15$ and $28$ are coprime, there is only one such location over the entire length of the cavity.

As expected, when the lattice field is detuned by an even number of FSRs from the probe field, the fluorescence is suppressed at the cavity center. For a detuning by an odd number of FSRs the suppression occurs half a period away from the cavity center.

The purely parabolic fits are performed including $\pm100$ pixels around the approximate cavity center position. From these we confirm that the cavity center is at an axial position of $320.70 \pm 0.15$ pixels. The center position for the individual measurements agree with the mean value within $+1.5$ and $-2.5$ pixels and their errorbars are within range of the mean. Converting the fit results into a physical length gives an uncertainty in the absolute positioning of the cavity center of only $\pm$135 nm, smaller than both the beating periods (here 400-700 $\mu$m) and the periodicity of the two standing waves (433 nm). As mentioned above, an even better precision could be obtained in principle by using more sets of detunings.

\section{Future prospects}
\label{Future}
As briefly mentioned in the introduction, the positioning of single or ensembles of ions with respect to the center of a Fabry-P\'{e}rot cavity may have several applications for ion-based cavity QED. First, it adds to the tools for trapping ions in localized optical fields~\cite{Schneider2010,Enderlein2012} for coherent atom-ion studies~\cite{Grier2009,Zipkes2010,Schmid2010,Cetina2012}. Second, superposing a steep and short-scale periodic optical potential to a shallow rf trap allows studies of structural~\cite{Drewsen2012,Horak2012,Cormick2012} and dynamical phase transitions (e.g. Coulomb-Frenkel-Kontorova model~\cite{GarciaMata2007,Benassi2011,Pruttivarasin2011,Schneider2012review,Cetina2013}). As demonstrated in~\cite{Linnet2012}, localizing an ion in a standing wave cavity field also allows for a better control of the ion-cavity coupling strength, which can be of interest for quantum information processing applications, such as single-photon generation~\cite{Keller2004,Barros2009}, quantum memory~\cite{Herskind2009,Albert2011}, photon counters~\cite{Clausen2013}, single ion-photon interfaces~\cite{Stute2012,Stute2013}, or for cavity-mediated cooling~\cite{Leibrandt2009}.

In addition to ionic systems in cavities, this positioning technique should also be applicable to e.g. cold neutral atomic species, trapped in magnetic or optical dipole traps. One can envision its application in single atom quantum dynamics studies~\cite{Pinkse2000,Hood2000,Thompson2013} which would benefit from accurate positioning. This include feedback~\cite{Kubanek2012}, cavity~\cite{Maunz2004,Nussmann2005} or ground-state cooling~\cite{Reiserer2013}. It also naturally applies to cavity QED studies with ensembles, e.g. to investigations of the quantum dynamics of cold atoms in cavity-generated optical potentials~\cite{Ritsch2013}, to applications involving the simultaneous interaction with multiple standing wave fields~\cite{Albert2011,Botter2013}, or to cold atom cavity optomechanics studies~\cite{StamperKurn2013}.

\section{Conclusion}
\label{Conclusion}
To conclude, we have demonstrated a simple method to accurately find the absolute center of a  Fabry-P\'{e}rot resonator using the spatially modulated fluorescence of trapped ions probed by two simultaneously resonating cavity fields. This method has potential applications to a wide range of cavity QED investigations with cold ions or atoms.

This work was supported by the European Commission (STREP
PICC and ITN CCQED) and the Carlsberg Foundation.


\begin{thebibliography}{100}

\bibitem{Paul1990} W. Paul, Rev. Mod. Phys. {\bf 62}, 531 (1990).

\bibitem{Grimm2000} R. Grimm, M. Weidem\"{u}ller, and Y. B. Ovchinnikov, Adv. At. Opt. Phys. {\bf 42}, 95 (2000).


\bibitem{BenDahan1996} M. BenDahan, E. Peik, J. Reichel, Y. Castin, and C. Salomon, Phys. Rev. Lett. {\bf 76}, 4508 (1996).

\bibitem{Greiner2002} M. Greiner, O. Mandel, T. Esslinger, T. W. H\"{a}nsch, and I. Bloch, Nature {\bf 415}, 39 (2002).

\bibitem{Wineland1975} D. J. Wineland and H. Dehmelt, Bull. Am. Phys. Soc.
{\bf 20}, 637 (1975).

\bibitem{Leibfried2003} D. Leibfried, R. Blatt, C. Monroe, and D. J. Wineland, Rev. Mod. Phys. {\bf 75}, 281 (2003).

\bibitem{Brown1986} L. S. Brown and G. Gabrielse, Rev. Mod. Phys. {\bf 58}, 233 (1986).

\bibitem{Dubin1999} D. H. E. Dubin and T. M. O'Neil, Rev. Mod. Phys. {\bf 71}, 87 (1999).

\bibitem{Blumel1988}  R. Blumel, J. M. Chen, E. Peik, W. Quint, W. Schleich,Y. R. Shen, and H. Walther, Nature {\bf 334}, 309 (2008).

\bibitem{Hoffnagle1988} J. Hoffnagle, R. G. DeVoe, L. Reyna, and R. G. Brewer. Phys. Rev. Lett. {\bf 61}, 255 (1988).

\bibitem{Blumel1989} R. Blumel, C. Kappler, W. Quint, and H. Walther, Phys. Rev. A {\bf 40}, 808 (1989).

\bibitem{Brewer1990} R. G. Brewer, J. Hoffnagle, and R. G. DeVoe, Phys. Rev. Lett. {\bf 65}, 2619 (1990).


\bibitem{Blatt2008} R. Blatt and D. J. Wineland, Nature {\bf 453}, 1008 (2008).

\bibitem{Willitsch2008} S. Willitsch, M. T. Bell, A. D. Gingell, and T. P. Softley, Phys. Chem. Chem. Phys. {\bf 10}, 7200 (2008).

\bibitem{Schneider2010} C. Schneider, M. Enderlein, T. Huber, S. D\"{u}rr, and T. Schaetz, Nature Phot. {\bf 4}, 772 (2010).

\bibitem{Enderlein2012} M. Enderlein, T. Huber, C. Schneider, and T. Schaetz, Phys. Rev. Lett. {\bf 109}, 233004 (2012).

\bibitem{Linnet2012} R. B. Linnet, I. D. Leroux, M. Marciante, A. Dantan, and M. Drewsen, Phys. Rev. Lett. {\bf 109}, 233005 (2012).

\bibitem{Karpa2013} L. Karpa, A. Bylinskii, D. Gangloff, M. Cetina, and V. Vuleti\'{c}, arxiv:1304.0049 (2013).

\bibitem{Cormick2011} C. Cormick, T. Schaetz, and G. Morigi, New J. Phys. {\bf 13}, 043019 (2011).

\bibitem{Drewsen2012} M. Drewsen, T. Matthey, A. Mortensen, and J. P. Hansen, arxiv:1202.2544 (2012).

\bibitem{Horak2012} P. Horak, A. Dantan, and M. Drewsen, Phys. Rev. A {\bf 86}, 043433 (2012).

\bibitem{Cormick2012} C. Cormick and G. Morigi, Phys. Rev. Lett. {\bf 109}, 053003 (2012).

\bibitem{GarciaMata2007} I. Garc\'{i}a- Mata, O. V. Zhirov, and D. L. Shepelyansky, Eur. J. Phys. D {\bf 41}, 325 (2007).

\bibitem{Benassi2011} A. Benassi, A. Vanossi, and E. Tosatti, Nat. Comm. {\bf 2}, 236 (2011).

\bibitem{Pruttivarasin2011} T. Pruttivarasin, M. Ramm, I. Talukdar, A. Kreuter, and H. H\"{a}ffner, New J. Phys. {\bf 13}, 075012 (2011).

\bibitem{Schneider2012review} C. Schneider, D. Porras, and T. Schaetz, Rep. Prog. Phys. {\bf 75}, 024401 (2012).

\bibitem{Herskind2009} P. F. Herskind, A. Dantan, J. P. Marler, M. Albert, and M. Drewsen, Nature Phys. {\bf 5}, 494 (2009).

\bibitem{Albert2011} M. Albert, A. Dantan, and M. Drewsen, Nature Phot. {\bf 5}, 633 (2011).

\bibitem{Clausen2013} C. Clausen, N. Sangouard, and M. Drewsen, New J. Phys. {\bf 15}, 025021 (2013).

\bibitem{Guthohrlein2001}  G. R. Guthohrlein, M. Keller, K. Hayasaka, W. Lange, and H. Walther, Nature {\bf 414}, 49 (2001).

\bibitem{Mundt2002} A. B. Mundt, A. Kreuter, C. Becher, D. Leibfried, J. Eschner, F. Schmidt-Kaler, and R. Blatt, Phys. Rev. Lett. {\bf 83}, 103001 (2002).

\bibitem{note} While the method is demonstrated here in the case of a symmetric Fabry-Perot cavity it could also be applied to a cavity with convex mirrors having different radii of curvature.

\bibitem{Herskind2009JPB} P. F. Herskind, A. Dantan, M. Albert, J. P. Marler, and M. Drewsen, J. Phys. B {\bf 42}, 154008 (2009).

\bibitem{Cohen-Tannoudji2004} C. Cohen-Tannoudji, J. Dupont-Roc, and G. Grynberg, \textit{Atom-photon interactions: basic processes and applications} (Wiley-VCH, Weinheim, 2004).

\bibitem{Herskind2008} P. F. Herskind, A. Dantan, M. B. Langkilde-Lauesen, A. Mortensen, J. L. S{\o}rensen, and M. Drewsen, Appl. Phys. B {\bf 93}, 373 (2008).

\bibitem{Grier2009} A. T. Grier, M. Cetina, F. Oru\v{c}evi\'{c}, and V. Vuleti\'{c}, Phys. Rev. Lett. {\bf 102}, 223201 (2009).

\bibitem{Zipkes2010} C. Zipkes, S. Palzer, L. Ratschbacher, C. Sias, and M. K\"{o}hl, Phys. Rev. Lett. {\bf 105}, 133201 (2010).

\bibitem{Schmid2010} S. Schmid, A. H\"{a}rter, and J. H. Denschlag, Phys. Rev. Lett. {\bf 105}, 133202 (2010).

\bibitem{Cetina2012} M. Cetina, A. T. Grier, and V. Vuleti\'{c}, Phys. Rev. Lett. {\bf 109}, 253201 (2012).

\bibitem{Cetina2013} M. Cetina, A. Bylinskii, L. Karpa, D. Gangloff, K. M. Beck, Y. Ge, M. Scholz, A. T. Grier, I. Chuang, and V. Vuleti\'{c}, arXiv:1302.2904 (2013).

\bibitem{Keller2004} M. Keller, B. Lange, K. Hayasaka, W. Lange, and H. Walther, Nature {\bf 431}, 1075 (2004).

\bibitem{Barros2009} H. G. Barros, A. Stute, T. E. Northup, C. Russo, P. O. Schmidt, and R. Blatt, New J. Phys. {\bf 11}, 103004 (2009).

\bibitem{Stute2012} A. Stute, B. Casabone, P. Schindler, T. Monz, P. O. Schmidt, B. Brandst\"{a}tter, T. E. Northup, and R. Blatt, Nature {\bf 485}, 482 (2012).

\bibitem{Stute2013} A. Stute, B. Casabone, B. Brandst\"{a}tter, K. Friebe, T. E. Northup, and R. Blatt, Nature Phot. {\bf 7}, 219 (2013).

\bibitem{Leibrandt2009} D. R. Leibrandt, J. Labaziewicz, V. Vuletic, and I. L. Chuang, Phys. Rev. Lett. {\bf 103}, 103001 (2009).


\bibitem{Pinkse2000} P. W. H. Pinkse, T. Fischer, P. Maunz, and G. Rempe, Nature {\bf 404}, 365 (2000).

\bibitem{Hood2000} C. J. Hood, T. W. Lynn, A. C. Doherty, A. S. Parkins, and H. J. Kimble, Science {\bf 287}, 1447 (2000).

\bibitem{Thompson2013}  J. D. Thompson, T. G. Tiecke, N. P. de Leon, J. Feist, A. V. Akimov, M. Gullans, A. S. Zibrov, V. Vuletic, and M. D. Lukin, Science {\bf 340}, 1202 (2013).

\bibitem{Maunz2004} P. Maunz, T. Puppe, I. Schuster, N. Syassen, P. W. H. Pinkse, and G. Rempe, Nature {\bf 428}, 50 (2004).

\bibitem{Nussmann2005} S. Nu{\ss}mann, K. Murr, M. Hijlkema, B. Weber, A. Kuhn, and G. Rempe, Nature Phys. {\bf 1}, 122 (2005).

\bibitem{Kubanek2012} A. Kubanek, M. Koch, C. Sames, A. Ourjoumtsev, P. W. H. Pinkse, K. Murr, and G. Rempe, Nature {\bf 462}, 898 (2012).

\bibitem{Reiserer2013} A. Reiserer, C. N\"{o}lleke, S. Ritter, and G. Rempe, Phys. Rev. Lett. {\bf 110}, 223003 (2013).

\bibitem{Ritsch2013} H. Ritsch, P. Domokos, F. Brennecke, and T. Esslinger, Rev. Mod. Phys. {\bf 85}, 553 (2013).

\bibitem{Botter2013} T. Botter, D. W. C. Brooks, S. Schreppler, N. Brahms, and D. M. Stamper-Kurn, Phys. Rev. Lett. {\bf 110}, 153001 (2013).

\bibitem{StamperKurn2013} D. M. Stamper-Kurn, arxiv:1204.4351 (2013).
\end{thebibliography}
\end{document}